\documentclass[reqno,11pt]{amsart}
\usepackage{bbm}
\usepackage{fullpage}
\usepackage{amsfonts}
\usepackage{amssymb}
\usepackage{graphicx}
\usepackage{arydshln}
\usepackage{latexsym}
\usepackage[mathscr]{eucal}
\usepackage{cite}
\usepackage{bm}
\usepackage{mathrsfs}
\usepackage{amsmath,amsthm,accents}
\numberwithin{equation}{section}
\DeclareMathOperator{\sech}{sech}

\usepackage{subfigure}
\numberwithin{equation}{section}

\newtheorem{Thm}{Theorem}
\def \wh#1{\widehat{#1}}
\def \wt#1{\widetilde{#1}}
\def \wb#1{\overline{#1}}

\def \dwh{\underaccent{{\cc@style\widehat{\mskip10mu}}}}
\def \dt#1{\underaccent{\tilde}{#1}}

\newcommand{\nn}{\nonumber}

\newcommand{\Ga}{\Gamma}

\newcommand{\al}{\alpha}
\newcommand{\be}{\beta}
\newcommand{\st}{\hbox{\tiny\it{T}}}

\begin{document}
	
\title{Solutions and continuum limits to nonlocal discrete modified Korteweg de-Vries equations}	
\author{Song-lin Zhao,~~Xiao-bo Xiang,~~Shou-feng Shen\\
\\\lowercase{\scshape{
Department of Applied Mathematics, Zhejiang University of Technology,
Hangzhou 310023, P.R. China}}}
\email{Corresponding Author: mathssf@zjut.edu.cn}

\begin{abstract}

In this paper, we take advantage of the bilinearization reduction method to consider the local and nonlocal reduction
of a discrete Ablowitz-Kaup-Newell-Segur equation.
Exact solutions in double Casoratian form to the reduced nonlocal discrete
modified Korteweg de-Vries equations are constructed. The dynamics
of soliton solutions are analyzed and illustrated by asymptotic analysis.
Moreover, both semi-continuous limit and full continuous limit,
are applied to obtain the local and nonlocal semi-discrete
modified Korteweg de-Vries equations, as well as the local and nonlocal continuous
modified Korteweg de-Vries equations.

\end{abstract}


\maketitle

\section{Introduction}

The modified Korteweg de-Vries (mKdV) equation
\begin{align}
\label{mKdV}
\al_t+\al_{xxx}+3\eta \al^2\al_x=0, \quad \eta=\pm 1
\end{align}
for $\al=\al(x,t)$ is completely integrable by means of the inverse scattering transform \cite{Wadati}, whose
general $N$-soliton solutions have been given by Hirota \cite{Hirota-mKdV}. Equation \eqref{mKdV}
adequately describes the long-wave approximation regime for cubic nonlinearity, where $\al$ is a
dimensionless electric field, and $x$ and $t$ are dimensionless space and time variables \cite{Leblond-1,Leblond-2}. This equation
arises in many physics contexts, such as
anharmonic lattices \cite{Ono}, Alf'ven waves \cite{Kakutani}, ion acoustic solitons \cite{Konno,Watanabe,Lonngren}, and so on.
Analogous to the most integrable equations, the mKdV equation and its complex form exhibit many interesting properties. Because of the discrete
reflection symmetry $\al\rightarrow -\al$
of the equation \eqref{mKdV}, there are two different types of soliton collisions, in which (up to reflection) the fast and
slow solitons in the collision have either the same orientation or
opposite orientations. The mKdV equation has breather solutions, which are not admitted by the KdV equation. In addition,
although there is no counterpart of KdV Galilean momentum for soliton solutions of the complex mKdV equation, they have an analogous Galilean
energy given by a conserved integral which is related to the motion of center of momentum (see \cite{Anco} and the references therein).

The mKdV equation can be deduced from the Ablowitz-Kaup-Newell-Segur (AKNS) system \cite{AKNS-1973}
\begin{subequations}
\label{AKNS}
\begin{align}
& \al_t+\al_{xxx}+3\al\be\al_x=0, \\
& \be_t+\be_{xxx}+3\al\be\be_x=0
\end{align}
\end{subequations}
with a standard symmetry reduction $\be(x,t)=\eta \al(x,t)$. If imposing nonlocal reduction
$\be(x,t)=\eta \al(-x, -t)$ or $\be(x,t)=\eta \al^*(-x, -t)$ on equation \eqref{AKNS}, then real/complex reverse space-time nonlocal mKdV
can be obtained \cite{AbMu-2016}, which are
\begin{subequations}
\label{rcst-mKdV}
\begin{align}
& \label{rst-mKdV}
\al_t+\al_{xxx}+3\eta \al\al(-x, -t)\al_x=0, \\
& \label{cst-mKdV}
\al_t+\al_{xxx}+3\eta \al\al^*(-x, -t)\al_x=0,
\end{align}
\end{subequations}
here and hereafter, the asterisk denotes complex conjugation.

Since the nonlocal nonlinear Schr\"{o}dinger equation with the parity-time
symmetry was introduced by Ablowitz and Musslimani in \cite{AbMu-2013},
it has been shown that nonlocal integrable systems have wide applications in mathematics and
physics. Mathematically, similar to local integrable systems, the nonlocal systems are always integrable, since they
have multi-soliton solutions, admit linear Lax pair formulation and possess an infinite number of conservation laws \cite{AM-Nonl}.
In physics, because of involvement of two different places $\{x, t\}$ and $\{x'=-x, t'=t\}$ (or $\{x'=-x, t'=-t\})$ in the equation,
the nonlocal integrable systems can be viewed as two-place systems \cite{Lou-SR}, which may be developed
to explain the correlated/entangled natural phenomena happened at two different spaces and/or times \cite{Lou-CTP}.
As typical integrable models, equations \eqref{rst-mKdV} and \eqref{cst-mKdV} have been actively studied from many different aspects, for example,
the exact solutions \cite{JZ-2017,Ji-Zhu,CDLZ,GP-CNSNS,Ma-JMAA,XZ-TMPH}, gauge equivalence \cite{Ma-GE},
long time asymptotics with finite density initial data \cite{Fan-PD},
transformation between the local mKdV equation and nonlocal mKdV equation \cite{YY-SAPM}, etc.

In the past few decades much direct attention have been paid to the (fully)
discrete integrable systems (see \cite{{N-2016-math}} and the references therein).
The discrete integrable systems mentioned here are the
integrable partial difference equations with discrete independent variables.
The research on discrete integrable systems not only brings new insights into the modern theory of integrable
systems, but also pushes forward the development of many subjects in pure mathematics, such
as algebraic geometry, Lie algebras, orthogonal polynomials, special functions and random matrices.
There exist many techniques to search for integrable discretization of differential equations,
among which a very important one is proposed by Hirota \cite{Hirota-1977} covering
a soliton preserving discretization of the direct (bilinear) method for nonlinear partial differential equations (see, e.g., \cite{Hirota-book}).
In \cite{Hirota-2000}, by discretizing the bilinear operators of AKNS equation \eqref{AKNS},
an integrable discrete AKNS (dAKNS) equation was introduced, namely,
\begin{subequations}
\label{dAKNS}
\begin{align}
& \wh{u}-u+\delta(1+uv)(\wt{u}-\wh{\dt{u}})w=0, \\
& \wh{v}-v+\delta(1+uv)(\wt{v}-\wh{\dt{v}})w=0, \\
& \label{dAKNS-c}
(1+\wh{u}\wh{v})\wt{w}=(1+uv)w,
\end{align}
\end{subequations}
which can transform back to the AKNS equation \eqref{AKNS}
under appropriate continuum limits. In \eqref{dAKNS} all dependent variables $u,~v$ and $w$ are functions defined on the two-dimensional
lattice with discrete coordinates $(n,m)\in \mathbb{Z}^2$, e.g., $u=u(n,m)$,
$\delta$ is a spacing parameter and tilde, hat serve as notations of shifts in different
directions
\begin{align}
\label{Nota-wth}
\wt{f}=f(n+1,m),\quad \wh{f}=f(n,m+1),
\end{align}
and naturally we have $\wh{\dt{f}}=f(n-1,m+1)$. Function $w$ here can be viewed as an auxiliary variable. From the
last equation \eqref{dAKNS-c}, one recognizes
\begin{align}
\label{dAKNS-c-rw}
w(n,m)=\prod_{j=n_0}^{n-1}\frac{1+{u}(j,m){v}(j,m)}{1+\wh u(j,m)\wh v(j,m)}, \quad n_0\in\mathbb{Z},
\end{align}
which possesses the nonautonomous structure (see Refs. \cite{SRH,GR}). Hirota in \cite{Hirota-2000}
also listed 1+1 soliton solution of equation \eqref{dAKNS}. Numerical integration
of the 1+1 soliton solution showed that soliton $u$ was damping and
$v$ was growing as time $m$ increases, which agreed exactly
with the theoretical ones except round-off errors.

Recent studies demonstrated how to extend the nonlocal theories
of continuous integrable systems to the semi-discrete integrable systems
\cite{SMMC,AM-2014,MZ-JMP,Ger,DLZ-AMC,FZS-IJMPB,ALM-Nonl,MSZ-AML}. Then
a natural question would be how to develop the classic method to study the
nonlocal discrete integrable equations. Despite the slow progress, some results have been achieved.
In \cite{ZKZ}, two types of nonlocal H1 equations were revealed, including
a reverse-$(n,m)$ nonlocal H1 equation and a reverse-$n$ nonlocal H1 equation. Recently,
Xiang et al. have systematically investigated exact solutions and continuum limits of the
nonlocal discrete sine-Gordon equations \cite{XFZ-TMPH,XZS-arXiv}, by utilizing the
Cauchy matrix reduction approach \cite{FZ-ROMP} (see also \cite{XZ-TMPH})
and the bilinearization reduction scheme \cite{CZ-AML,LWZ-SAPM}, respectively. Compared with the
Cauchy matrix reduction approach, the bilinearization reduction scheme has its own preponderance,
since it can be used to the survey of real nonlocal integrable systems. Moreover, the soliton-Jordan
block mixed solutions can not be generated in the Cauchy matrix reduction framework.

In this paper, we would like to study nonlocal reductions of the dAKNS equation \eqref{dAKNS} from
the perspective of bilinearization reduction. In this issue, a real nonlocal discrete mKdV equation and a
complex nonlocal discrete mKdV equation will be shown. Through solving the matrix equations,
soliton solutions and Jordan block solutions in the form of double Casorati determinants will be
exhibited. In particular, we will focus on the one-soliton, two-solitons and the simplest Jordan block solutions, as well as
their dynamics. Continuum limits for the resulting nonlocal discrete mKdV equations will be also
discussed.

In Section 2, we present some preliminaries and construct double Casoratian solutions to the dAKNS equation \eqref{dAKNS}.
Then we consider in Section 3 real local and nonlocal reduction of the
dAKNS equation \eqref{dAKNS}. Some types of exact solutions and their dynamics
for the resulting real nonlocal discrete mKdV equation are derived. Furthermore, continuum limits are
introduced to discuss the real local and nonlocal semi-discrete mKdV equation and
the real local and nonlocal continuous mKdV equation. In Section 4 we focus on the complex local
and nonlocal reduction of the dAKNS equation \eqref{dAKNS}. Finally,
in the last section we give some conclusions about the obtained results.

\section{Double Casoratian solutions to the dAKNS equation \eqref{dAKNS}}

Given the two basic column vectors
\[\Phi=(\Phi_1, \Phi_2,\cdots,\Phi_{N+M+2})^{\st}, \quad \Psi=(\Psi_1, \Psi_2,\cdots,\Psi_{N+M+2})^{\st}\]
with $\Phi_j:=\Phi_j(n,m)$ and $\Psi_j:=\Psi_j(n,m)$, $(j=1,2,\ldots,N+M+2)$, we are interested in the condition equation set (CES)
\begin{subequations}
\label{dAKNS-CES}
\begin{align}
& \wt{\Phi}=e^K\Phi,\quad \wh{\Phi}=[(\delta E^{2}-1)
(\delta E^{-2}-1)^{-1}]^{\frac{1}{2}}\Phi,\\
& \wt{\Psi}=e^{-K}\Psi,\quad
\wh{\Psi}=[(\delta E^{2}-1)
(\delta E^{-2}-1)^{-1}]^{\frac{1}{2}}\Psi,
\end{align}
\end{subequations}
where $K$ is an invertible constant matrix of order $N+M+2$ and
$E$ is a shift operator defined as $E^jf(n,m)=f(n+j,m)$, ($j \in \mathbb{Z}$).
For the vectors $\Phi$ and $\Psi$ satisfying CES \eqref{dAKNS-CES}, we represent
the double Casoratian as
\begin{align}
	\mbox{C}^{(N,M)}(\Phi, \Psi)& =|\Phi, E^2\Phi, \ldots, E^{2N}\Phi; \Psi, E^2\Psi, \ldots, E^{2M}\Psi| \nn \\
	& =|0, 1,\ldots, N;0,1,\ldots, M|=|\wh{N};\wh{M}|.
\end{align}
In terms of the Freeman-Nimmo's shorthand notation (cf.\cite{FN-1983}), we have compactly written
the set of consecutive columns $0,1,\cdots,N$ as $\wh{N}$,
which cannot be confused with the use of hat for shifts.


\subsection{Bilinearization and double Casoratian solutions}

Through the transformation of dependent variables
\begin{align}
\label{dAKNS-uvw-tran}
u=g/f,\quad v=h/f,\quad w=f\wh{\dt f}/(\wh{f}\dt{f}),
\end{align}
the equation \eqref{dAKNS} can be transformed into
\begin{subequations}\label{dAKNS-bili}
\begin{align}
& \wh{g}f-g\wh{f}+\delta(\wt{g}\wh{\dt f}-\wh{\dt g}\wt{f})=0, \\
& \wh{h}f-h\wh{f}+\delta(\wt{h}\wh{\dt f}-\wh{\dt g}\wt{f})=0, \\	
& \wt{f}\dt{f}-f^2=gh.
\end{align}
\end{subequations}
It is readily to see that this bilinear system is invariant under gauge transformation
\begin{align}
f\rightarrow f~\mbox{exp}(a n+b m), \quad
g\rightarrow g~\mbox{exp}(a n+b m), \quad
h\rightarrow h~\mbox{exp}(a n+b m),
\end{align}
where $a$ and $b$ are two constants.

The double Casoratian solutions of the bilinear system \eqref{dAKNS-bili} are listed in the following theorem.
\begin{Thm}
The double Casorati determinants
\begin{align}
\label{dAKNS-dCs}
f=|\wh{N};\wh{M}|,\quad g=|\wh{N+1};\wh{M-1}|, \quad h= |\wh{N-1};\wh{M+1}|
\end{align}	
solve the bilinear system \eqref{dAKNS-bili}, provided that the basic column vectors $\Phi$ and
$\Psi$ are given by the CES \eqref{dAKNS-CES}.
\end{Thm}

A straightforward calculation of CES \eqref{dAKNS-CES} yields
\begin{subequations}
\label{dAKNS-Phsi-K}
\begin{align}
&\Phi=e^{Kn}[(\delta e^{2K}-I)(\delta e^{-2K}-I)^{-1}]^{\frac{m}{2}}C^{+}, \\
&\Psi=e^{-Kn}[(\delta e^{2K}-I)(\delta e^{-2K}-I)^{-1}]^{-\frac{m}{2}}C^{-},
\end{align}
\end{subequations}
where $C^{\pm}=(c_{1}^{\pm},c_{2}^{\pm},\dots c_{N+1}^{\pm};d_{1}^{\pm},d_{2}^{\pm},\dots d_{M+1}^{\pm})^{\st}$ are constant column vectors.
Here and hereafter $I$ is the unit matrix whose index indicating the size is omitted.

\subsection{Similarity invariance of exact solutions}

We denote \eqref{dAKNS-dCs}
composed by \eqref{dAKNS-Phsi-K} by $f(\Phi,\Psi)$, $g(\Phi,\Psi)$ and $h(\Phi,\Psi)$.
To illustrate the similarity invariance of exact solutions \eqref{dAKNS-uvw-tran}, we introduce a matrix $\wb{K}$ that is similar to $K$, i.e.,
\begin{align}
\label{Ga-K}
\wb{K}=TKT^{-1},
\end{align}
where $T$ is the transform matrix. By substituting \eqref{Ga-K} into \eqref{dAKNS-Phsi-K} and denoting
$\wb{C}^{\pm}=TC^{\pm}$, we conclude that the new basic column vectors
\begin{subequations}
\label{dKANS-Phsi-wbK}
\begin{align}
&\wb{\Phi}=T\Phi=e^{\wb{K} n}[(\delta e^{2\wb{K}}-1)(\delta e^{-2\wb{K}}-1)^{-1}]^{\frac{m}{2}}\wb{C}^{+}, \\
&\wb{\Psi}=T\Psi=e^{-\wb{K} n}[(\delta e^{2\wb{K}}-1)(\delta e^{-2\wb{K}}-1)^{-1}]^{-\frac{m}{2}}\wb{C}^{-}
\end{align}
\end{subequations}
still satisfy the CES \eqref{dAKNS-CES} with $K\rightarrow \wb{K}$.
Noticing the relations $f(\wb{\Phi},\wb{\Psi})=|T|f(\Phi,\Psi)$, $g(\wb{\Phi},\wb{\Psi})=|T|g(\Phi,\Psi)$ and $h(\wb{\Phi},\wb{\Psi})=|T|h(\Phi,\Psi)$,
one can easily know that $(f(\wb{\Phi},\wb{\Psi}),g(\wb{\Phi},\wb{\Psi}),h(\wb{\Phi},\wb{\Psi}))$ and
$(f(\Phi,\Psi),g(\Phi,\Psi),h(\Phi,\Psi))$ lead to same solutions for the
dAKNS equation \eqref{dAKNS} through transformation \eqref{dAKNS-uvw-tran}.
Thus it is feasible to let $K$ being its Jordan canonical form $\Ga$ in \eqref{dAKNS-Phsi-K}, i.e.,
\begin{subequations}
\label{dAKNS-Phsi-Ga}
\begin{align}
& \label{dAKNS-Phi-Ga}
\Phi=e^{\Ga n}[(\delta e^{2\Ga}-I)(\delta e^{-2\Ga}-I)^{-1}]^{\frac{m}{2}}C^{+}, \\
&\Psi=e^{-\Ga n}[(\delta e^{2\Ga}-I)(\delta e^{-2\Ga}-I)^{-1}]^{-\frac{m}{2}}C^{-}.
\end{align}
\end{subequations}

In the following two sections, our first task is to show the bilinearization reduction
framework for the dAKNS equation \eqref{dAKNS} in the real sense, respectively, complex sense.
This scheme involves, first taking
appropriate reductions to get the local and nonlocal discrete mKdV equations, and second solving the matrix
equation algebraically to derive the exact solutions. To construct
explicit solutions of the reduced mKdV equations,
we take $M=N$ and replace $C^{\pm}$ by $e^{\mp N\Gamma}C^{\pm}$ in \eqref{dAKNS-Phsi-Ga} and
pay attention to the following double Casorati determinants:
\begin{align*}
f=|e^{-N\Gamma}\wh{\Phi^{(N)}};e^{N\Gamma}\wh{\Psi^{(N)}}|,\;
g= |e^{-N\Gamma}\wh{\Phi^{(N+1)}};e^{N\Gamma}\wh{\Psi^{(N-1)}}|,\;
h=|e^{-N\Gamma}\wh{\Phi^{(N-1)}};e^{N\Gamma}\wh{\Psi^{(N+1)}}|.
\end{align*}
Next, we will inspect continuum limits of the resulting
local and nonlocal discrete mKdV equations, as well as exact solutions and dynamics.
In what follows, for the function $\mathbb{F}:=\mathbb{F}(x_1,x_2)$ we appoint that notation
$\mathbb{F}_\sigma$ means $\mathbb{F}_\sigma=\mathbb{F}(\sigma x_1, \sigma x_2)$, where $\sigma=\pm 1$ (cf. \cite{XZS-arXiv}).
The notation $\mathbb{F}_{\sigma}=\mathbb{F}$ for $\sigma=1$ should not be confused
with the component $\mathbb{F}_1$ in matrix.

\section{Real reductions: solutions and continuum limits}


We begin with the constraint
\begin{align}
\label{real-Re}
v=\eta u_\sigma, \quad \eta, \sigma=\pm 1.
\end{align}	
Substituting \eqref{real-Re} into dAKNS \eqref{dAKNS} and noting that $w=w_{\sigma}$, we catch the real
local and nonlocal discrete mKdV (rnd-mKdV) equation
\begin{subequations}
\label{rndmKdV}
\begin{align}
& \widehat{u}-u+\delta(1+\eta u u_\sigma)(\widetilde{u}-\widehat{\underaccent{\tilde}{u}})w=0, \\
\label{rnd-mKdV-rb}
& (1+\eta\wh{u}\wh{u}_{\sigma})\wt w=(1+\eta uu_{\sigma})w.
\end{align}
\end{subequations}
When $\sigma=1$, equation \eqref{rndmKdV} is the real local discrete mKdV equation, while when
$\sigma=-1$, equation \eqref{rndmKdV} is the real nonlocal discrete mKdV equation. It was checked that
equation \eqref{rndmKdV} is preserved under transformation $u\rightarrow -u$. Besides, equation \eqref{rndmKdV} with
$(\sigma,\eta)=(\pm 1,1)$ and with $(\sigma,\eta)=(\pm 1,-1)$ can be
transformed into each other by taking $u\rightarrow iu$.

For the double Casoratian solution of equation \eqref{rndmKdV},
we show the results in the following theorem.
\begin{Thm}
The functions $u=g/f$ and $w=f\wh{\dt f}/(\wh{f}\dt{f})$ with
\begin{align}
\label{dmKdV-solu}
f=|e^{-N\Gamma}\wh{\Phi^{(N)}};e^{N\Gamma}\wh{\Psi^{(N)}}|,\quad
g=|e^{-N\Gamma}\wh{\Phi^{(N+1)}};e^{N\Gamma}\wh{\Psi^{(N-1)}}|
\end{align}
solve the rnd-mKdV equation \eqref{rndmKdV}, if the $(2N+2)$-th order column vectors
$\Phi$ and $\Psi$ defined by \eqref{dAKNS-Phsi-Ga} satisfy the relation
\begin{align}
\label{dmKdV-Phsi-T}
\Psi=T\Phi_\sigma,
\end{align}
where $T\in\mathbb{C}^{(2N+2)\times(2N+2)}$ is a constant matrix satisfying
\begin{align}
\label{rndmKdV-AT}
\Gamma T+\sigma T\Gamma=\bm 0,\quad T^{2}=\left\{
\begin{array}{l}
-\eta I, \quad \mbox{with} \quad \sigma=1,\\
\eta|e^\Gamma|^{2}I, \quad \mbox{with} \quad \sigma=-1,
\end{array}\right.
\end{align}
and we require $C^{-}=TC^{+}$.
\end{Thm}
For the proof of Theorem 2, one can refer to a similar one given in \cite{XZS-arXiv}. We call the two equations
in \eqref{rndmKdV-AT} as matrix equations, in which the first one is the famous Sylvester equation (cf. \cite{Syl,BR}),
usually appearing in systems and control theory, signal processing, filtering, model reduction,
image restoration, and so on. Based on this theorem, we aware that
$u=g/f$ and $w=f\wh{\dt f}/(\wh{f}\dt{f})$ with
\begin{align}
f=|e^{-N\Gamma}\wh{\Phi^{(N)}};e^{N\Gamma}T\wh{\Phi_{\sigma}^{(-N)}}|,\quad
g=|e^{-N\Gamma}\wh{\Phi^{(N+1)}};e^{N\Gamma}T\wh{\Phi_{\sigma}^{(-N+1)}}|
\end{align}
solve the rnd-mKdV equation \eqref{rndmKdV}, where $T$ and $\Gamma$ satisfy the matrix equations
\eqref{rndmKdV-AT}.

In what follows, we will solve the matrix equations \eqref{rndmKdV-AT}
to display some exact solutions of the rnd-mKdV equation \eqref{rndmKdV},
comprising soliton solutions and Jordan block solutions. We just consider the nonlocal case, i.e., $\sigma=-1$.
Without loss of generality, we set $\eta=1$. For the sake of brevity, we introduce some notations
\begin{align}
\label{xi}
\xi_j=k_jn+\tau_j m, \quad e^{\tau_j}=\bigg(
\dfrac{\delta e^{2k_{j}}-1}{\delta e^{-2k_{j}}-1}\bigg)^{\frac{1}{2}}, \quad j=1,2,\ldots,N+1,
\end{align}	
where we assume $\delta^2+1>2\delta\cosh 2k_{j}$ to guarantee the real property of $\tau_j$.

\subsection{Some examples of solutions}

In the case of $(\eta=1,\sigma=-1)$, solution for equation \eqref{rndmKdV-AT} is taken as
\begin{align}
	\label{Ga-T-ex}
	\Ga=\left(
	\begin{array}{cc}
		L & \bm 0  \\
		\bm 0 & -L
	\end{array}\right),\quad
	T=\left(
	\begin{array}{cc}
		I & \bm 0  \\
		\bm 0 & -I
	\end{array}\right),
\end{align}	
where $L$ is the Jordan canonical matrix. The exact solutions, either mult-soliton solutions or
Jordan block solutions, rest with the eigenvalue structure of matrix $L$. More explicitly,
when $L$ is a diagonal matrix composed by distinct nonzero eigenvalues, soliton solutions can be derived. While when $L$ is a Jordan block matrix,
Jordan block solutions can be obtained. In general, the Jordan block solutions can be generated from
soliton solutions through a special limit procedure (see Refs. \cite{AS-JMP,ZDJ-Wron,ZZSZ}).
Due to the block structure of matrix $T$, we note that $C^{+}$ can be
gauged to be $\bar{I}=(1,1,\dots,1;1,1,\dots,1)^{\st}$ or $\breve{I}=(1,0,\dots,0;1,0,\dots,0)^{\st}$.
This implies that the solutions we obtain for the equation \eqref{rndmKdV} are independent of phase parameters $C^{+}$,
i.e., the initial phase has always to be 0.

\vspace{.2cm}
\noindent{\it Soliton solutions}: When $L$ is a diagonal matrix
\begin{align}
\label{drL-Diag}
L=\mathrm{Diag}(k_{1},k_{2},\dots,k_{N+1}), \quad |L| \neq 0, \quad k_i \neq k_j, \quad (i \neq j),
\end{align}	
then we take $C^{+}=\bar{I}$ and have
\begin{align*}
\Phi_{j}=\left\{
\begin{aligned}
& e^{\xi_j},\quad j=1, 2, \ldots, N+1, \\
& e^{-\xi_{s}},\quad j=N+1+s,\quad s=1,2,\ldots,N+1.
\end{aligned}	
\right.
\end{align*}

Order $N=0$ yields one-soliton solution
\begin{subequations}
\label{dr-1ss}
\begin{align}
\label{dr-1ss-u}
& u=\sinh 2k_{1}\sech 2\xi_1, \\
& w=\cosh2\xi_1\cosh2(\tau_1+\xi_1-k_{1})\sech2(\xi_1+\tau_1)\sech2(\xi_1-k_{1}).
\end{align}
\end{subequations}
Solution $u$ in \eqref{dr-1ss-u} describes a single peaked and unidirectional travelling
wave, whose top trace and amplitude are, respectively, $n=-\tau_1m/k_1$ and $\sinh 2k_{1}$.
The width is proportional to $(2k_1)^{-1}$ and the speed is $-\tau_1/k_1$. In terms of the sign of the parameter $k_1$, the amplitude of the wave
can be positive or negative, which corresponds to soliton or anti-soliton, as depicted in Fig. 1.
\vskip20pt
\begin{center}
	\begin{picture}(120,80)
		\put(-170,-23){\resizebox{!}{4cm}{\includegraphics{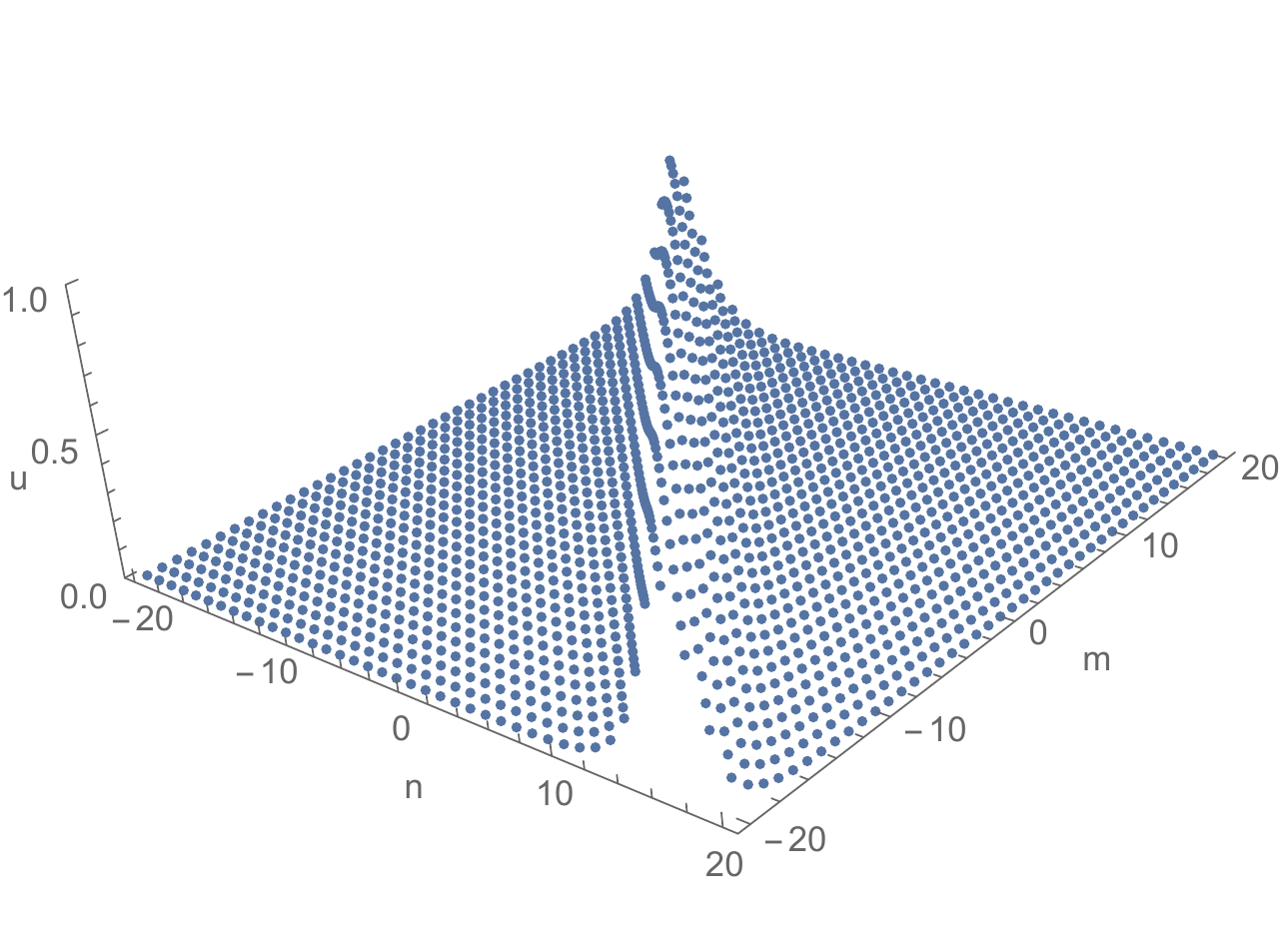}}}
		\put(10,-23){\resizebox{!}{3.5cm}{\includegraphics{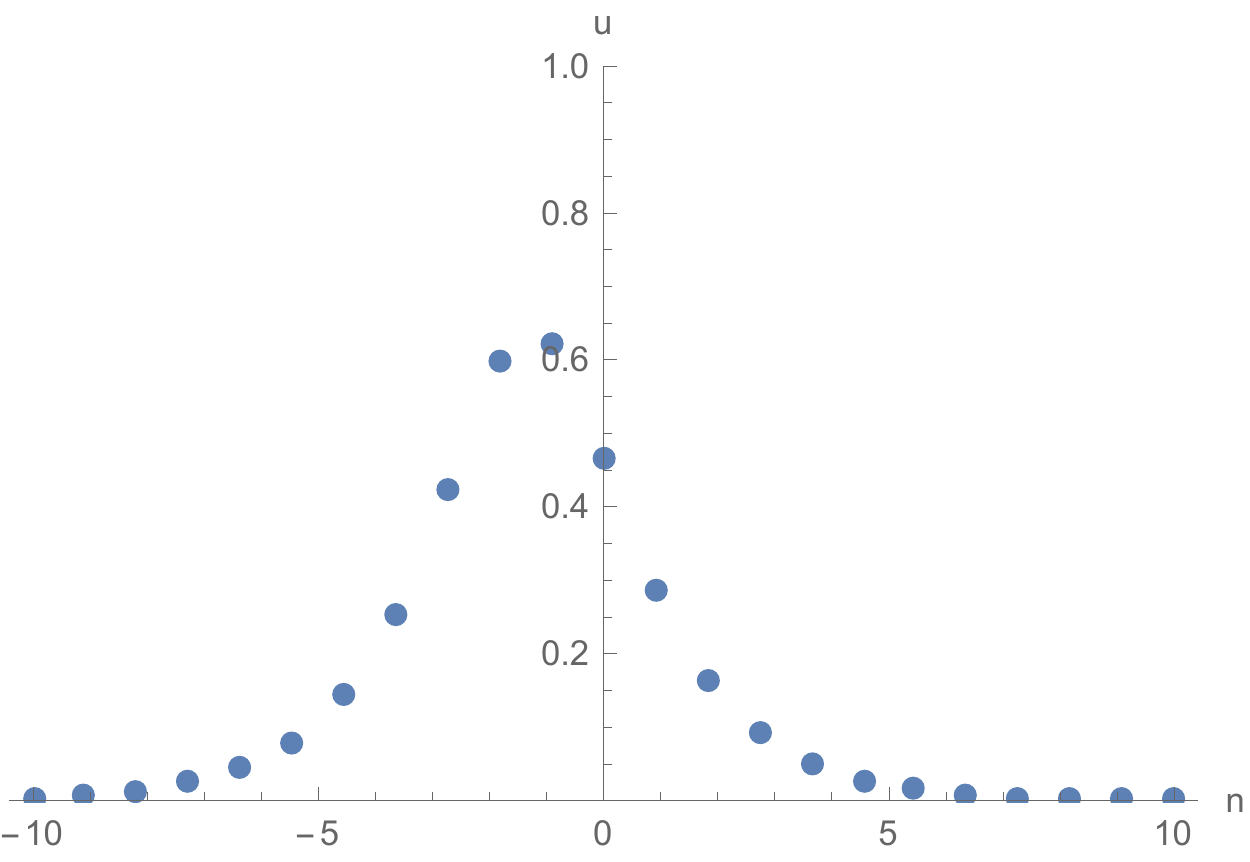}}}
		\put(150,-23){\resizebox{!}{3.5cm}{\includegraphics{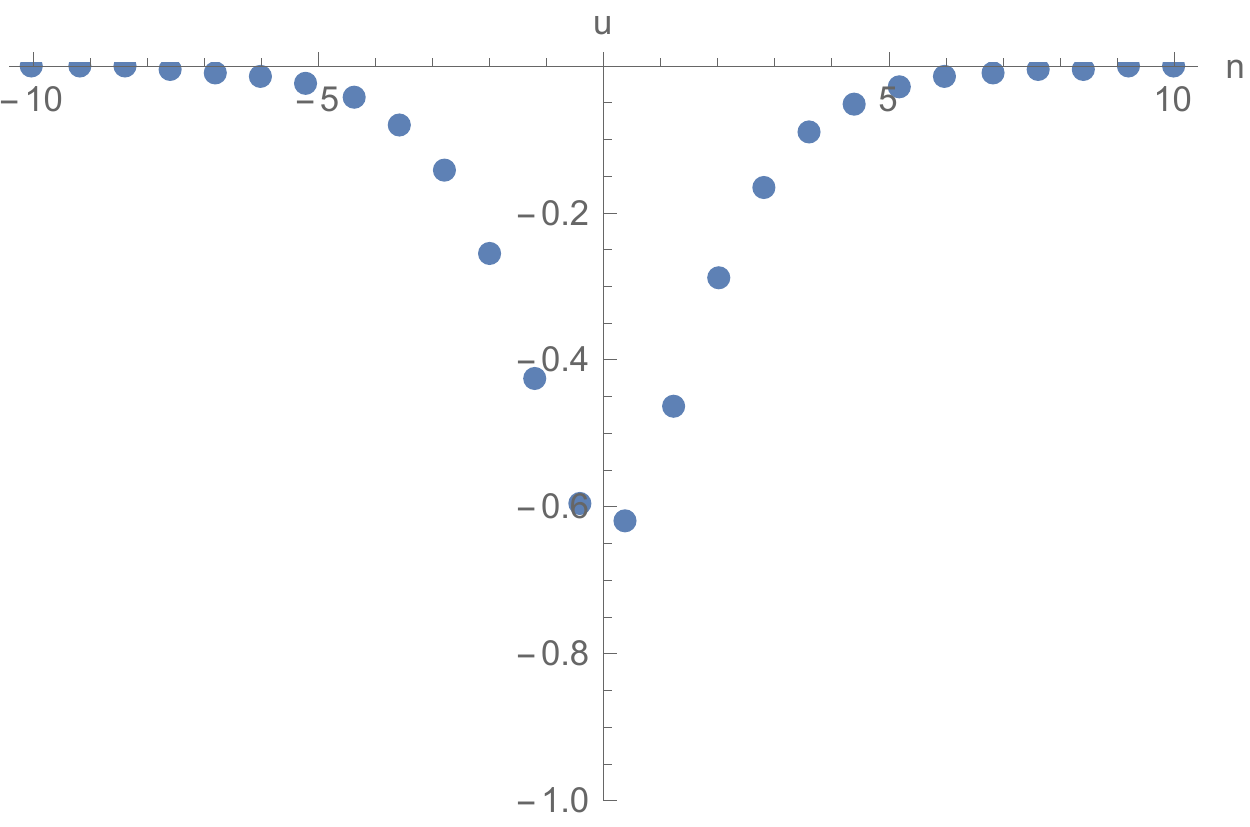}}}
	\end{picture}
\end{center}
\vskip 30pt
\begin{center}
\begin{minipage}{15cm}{\footnotesize
\qquad\qquad\qquad\quad(a)\qquad\qquad\qquad\qquad\qquad\qquad\qquad\qquad (b) \qquad\qquad\qquad\qquad\qquad\qquad\quad\quad (c)\\
{\bf Fig. 1.} One-soliton solution $u$ given by \eqref{dr-1ss-u} with $\delta=-1.5$:
(a) shape and movement for $k_1=0.5$;
(b) soliton for $m=2$ and $k_1=0.5$;
(c) anti-soliton for $m=2$ and $k_1=-0.5$.}
\end{minipage}
\end{center}

Order $N=1$ leads to the two-soliton solutions
\begin{align}
\label{dr-2ss}
u=g/f, \quad w=f\wh{\dt f}/(\wh{f}\dt{f}),
\end{align}	
in which
\begin{subequations}
\begin{align*}
f=& \cosh2(k_{1}+k_{2})\sinh^2(\xi_1-\xi_2)+\cosh2(k_{1}-k_{2})\cosh^2(\xi_1+\xi_2)-\cosh2\xi_1\cosh2\xi_2, \\
g=& (\cosh2k_{1}-\cosh2k_{2})(\sinh2k_{1}\cosh2\xi_{2}-\sinh2k_{2}\cosh 2\xi_{1}).
\end{align*}	
\end{subequations}
With the help of the analysis of moving-coordinate expansions \cite{Hie-book}, we can identify the
asymptotic form of the two-soliton solutions $u$ given in \eqref{dr-2ss} as $m\rightarrow \pm \infty$.
Without the loss of generality, we set $\delta<-1$ and $0<k_{1}<k_{2}$ to guarantee $\tau_1>\tau_2>0$.
We keep $\xi_1$ constant and let $m$ go to infinity, and find that there is only $k_1$-soliton left
along the line $\xi_1$=const. Moreover, we also find how the asymptotic $k_1$-soliton is
distinguished by its top traces, amplitude and speed.  After similar
and also detailed discussions for the $k_2$-soliton, we can make a clear description on
two-soliton interactions.
\begin{Thm}
Suppose $\delta<-1$ and $0<k_{1}<k_{2}$. Then, when $m\rightarrow \pm \infty$, the $k_1$-soliton asymptotically follows
\begin{subequations}
\begin{align}
top \: point \: traces&: n(m)=\pm \dfrac{1}{2k_{1}}\ln\dfrac{\sinh(k_{2}-k_{1})}
{\sinh(k_{1}+k_{2})}-\frac{\tau_1}{k_1}m, \\
amplitude &: u=\dfrac{\sinh2k_{1}(\cosh2k_{1}-\cosh2k_{2})}
{2\sinh(k_{1}+k_{2})\sinh(k_{2}-k_{1})}, \\
speed &: -\frac{\tau_1}{k_1}, \\
phase \: shift &: \dfrac{1}{k_{1}}\ln\dfrac{\sinh(k_{2}-k_{1})}
{\sinh(k_{1}+k_{2})},
\end{align}	
\end{subequations}
and the $k_2$-soliton asymptotically follows
\begin{subequations}
\begin{align}
top \: point \: traces&: n(m)=\mp \dfrac{1}{2k_{2}}\ln\dfrac{\sinh(k_{2}-k_{1})}
{\sinh(k_{1}+k_{2})}-\frac{\tau_2}{k_2}m,\\
amplitude&: u=\dfrac{\sinh2k_{2}(\cosh2k_{2}-\cosh2k_{1})}
{2\sinh(k_{1}+k_{2})\sinh(k_{2}-k_{1})},\\
speed&: -\frac{\tau_2}{k_2}, \\
phase \: shift &: -\dfrac{1}{k_{2}}\ln\dfrac{\sinh(k_{2}-k_{1})}
{\sinh(k_{1}+k_{2})}.
\end{align}	
\end{subequations}
\end{Thm}
The solution $u$ given in \eqref{dr-2ss} is illustrated in Fig. 2.
\vskip20pt
\begin{center}
	\begin{picture}(120,80)
	\put(-170,-23){\resizebox{!}{3.5cm}{\includegraphics{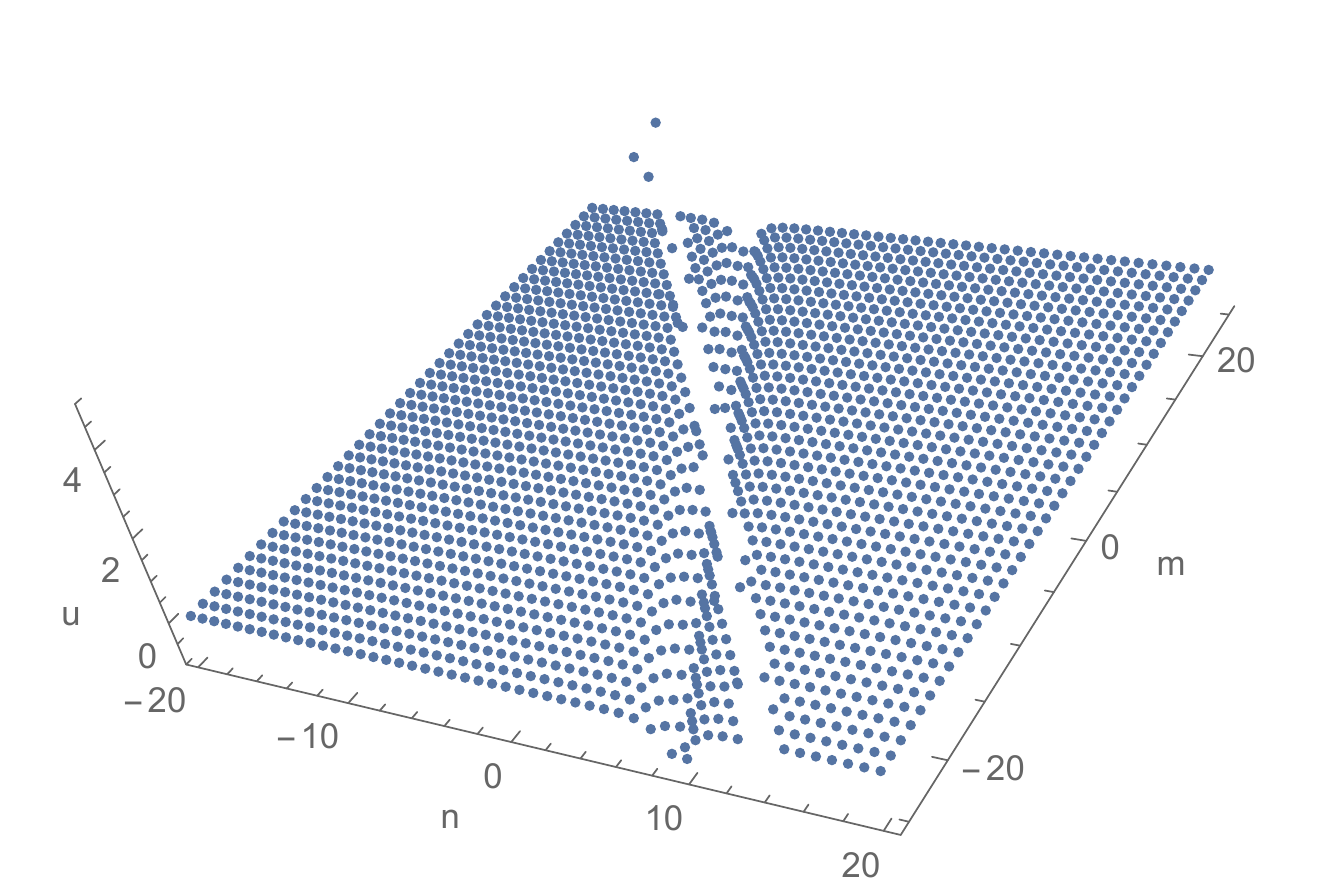}}}
	\put(-10,-23){\resizebox{!}{3.5cm}{\includegraphics{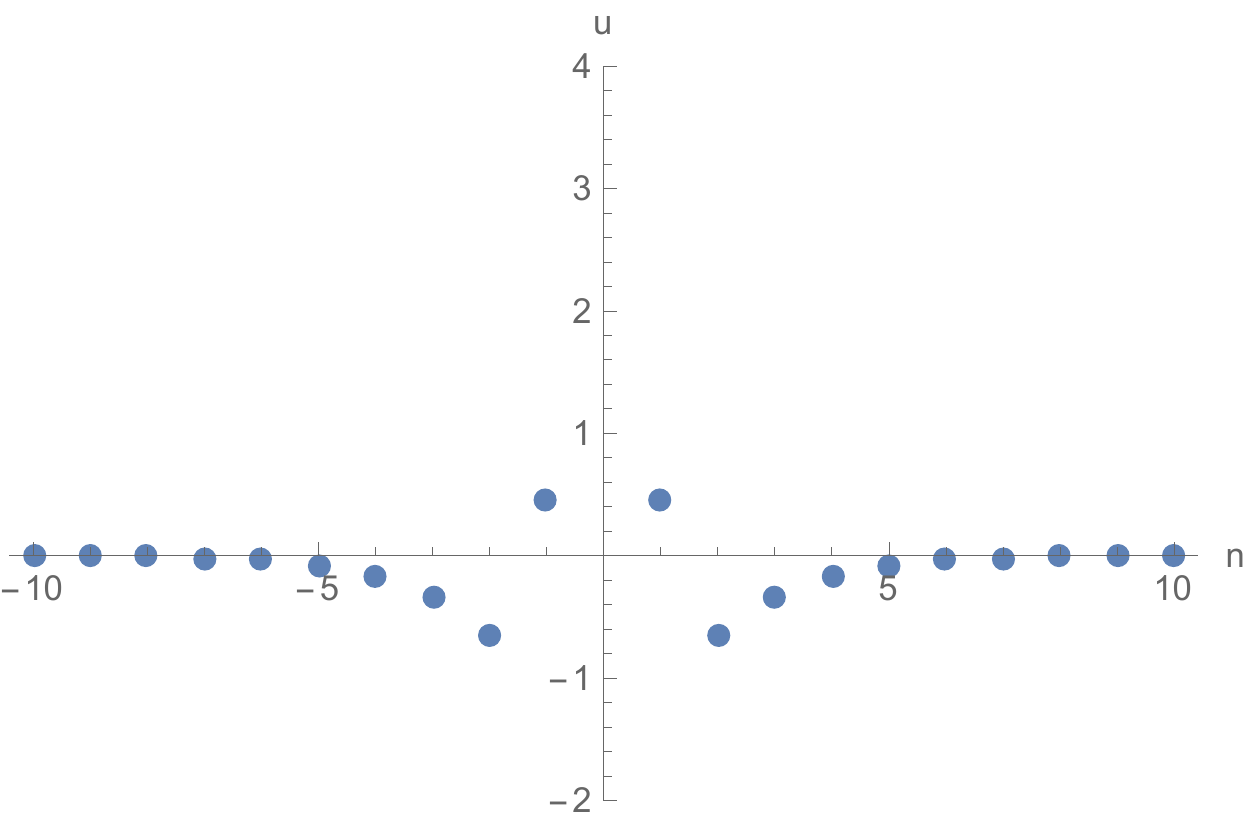}}}
	\put(150,-23){\resizebox{!}{3.5cm}{\includegraphics{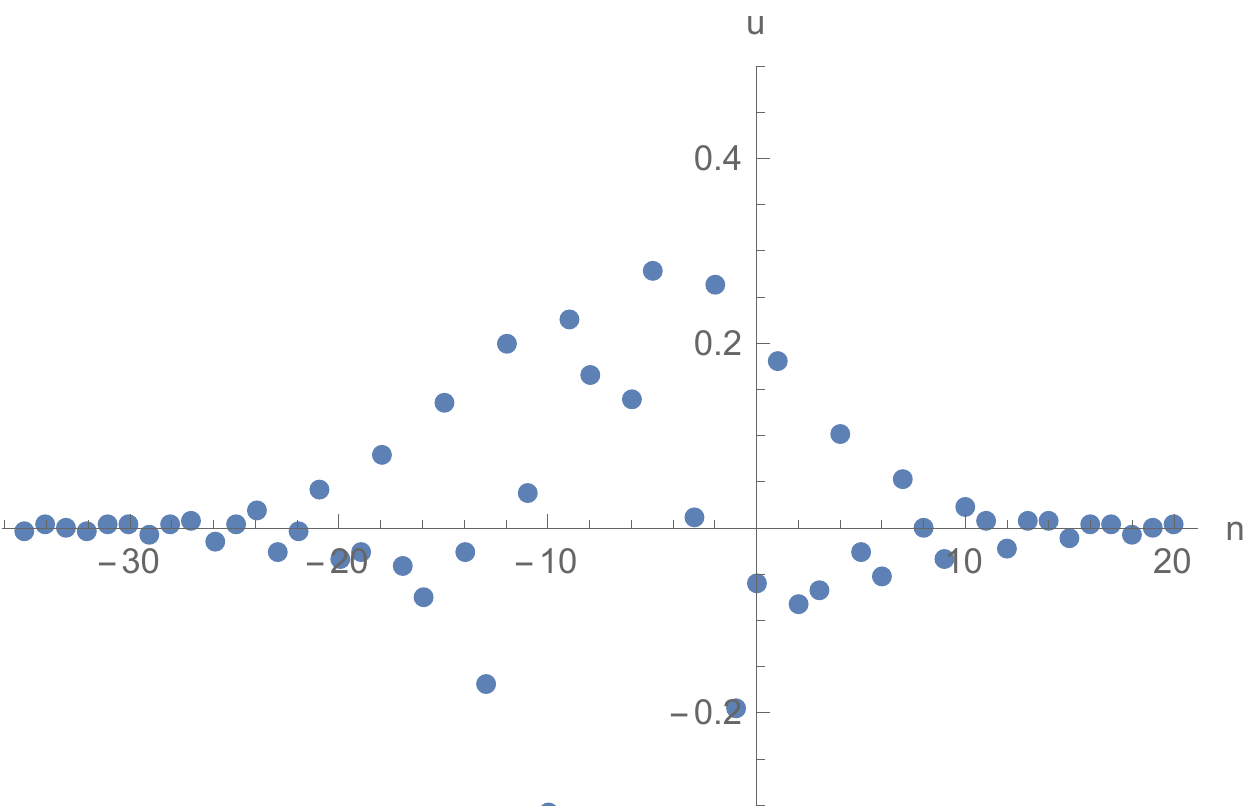}}}
	\end{picture}
\end{center}
\vskip 20pt
\begin{center}
	\begin{minipage}{15cm}{\footnotesize
			\qquad\qquad\qquad\quad(a)\qquad\qquad\qquad\qquad\qquad\qquad\qquad\qquad (b) \qquad\qquad\qquad\qquad\qquad\qquad\quad\qquad (c)\\
			{\bf Fig. 2.} Two-soliton solutions $u$ given by \eqref{dr-2ss} with $\delta=-5$:
			(a) shape and motion with $k_1=0.4$ and $k_2=6$; (b) 2D-plot of (a) at $m=0$; (c) breather solution with $k_1=0.1+i$ and $k_2=0.1-i$
			at $m=1$.}
	\end{minipage}
\end{center}
\vspace{.2cm}

\noindent{\it Jordan block solutions}:
When $L$ is a Jordan block matrix
\begin{align}
\label{drL-Jor}
L=\left(
\begin{array}{cccc}
k_1 &   0 &  \cdots&  0\\
1 & k_1 &  \cdots&  0\\		
\vdots& \ddots&\ddots&\vdots\\
0&\cdots& 1 & k_1
\end{array}\right)_{(N+1)\times(N+1)},\quad k_1 \neq 0,
\end{align}	
we take $C^{+}=\breve{I}$ and get
\begin{align*}
	\Phi_{j}=\left\{
	\begin{aligned}
		&\dfrac{\partial^{j-1}_{k_1}e^{\xi_1}}{(j-1)!}, \quad j=1,2, \dots, N+1,\\
		&\dfrac{\partial^{s-1}_{k_1}e^{-\xi_1}}{(s-1)!}, \quad j=N+1+s,\quad s=1,2,\dots,N+1.
	\end{aligned}	
	\right.
\end{align*}

In the case of $N=1$, we write down the simplest Jordan block solution
\begin{subequations}
	\begin{align}
		\label{dr-JB-solu-u}
		& u=\dfrac{\cosh2\xi_{1}\sinh4k_{1}-2\varsigma\sinh2\xi_{1}\sinh^{2}2k_{1}}
		{\varsigma^{2}\sinh^{2}2k_{1}+\cosh^{2}2\xi_{1}}, \\
		& w=f\wh{\dt f}/(\wh{f}\dt{f}), \quad f=\varsigma^{2}\sinh^{2}2k_{1}+\cosh^{2}2\xi_{1},
	\end{align}	
\end{subequations}
where $\varsigma=n+m\bigg(\dfrac{e^{2k_{1}}-2\delta}{e^{2k_{1}}-\delta}+\dfrac{1}{\delta e^{2k_{1}}-1}\bigg)$.
\vskip20pt
\begin{center}
	\begin{picture}(120,80)
		\put(-140,-23){\resizebox{!}{4.0cm}{\includegraphics{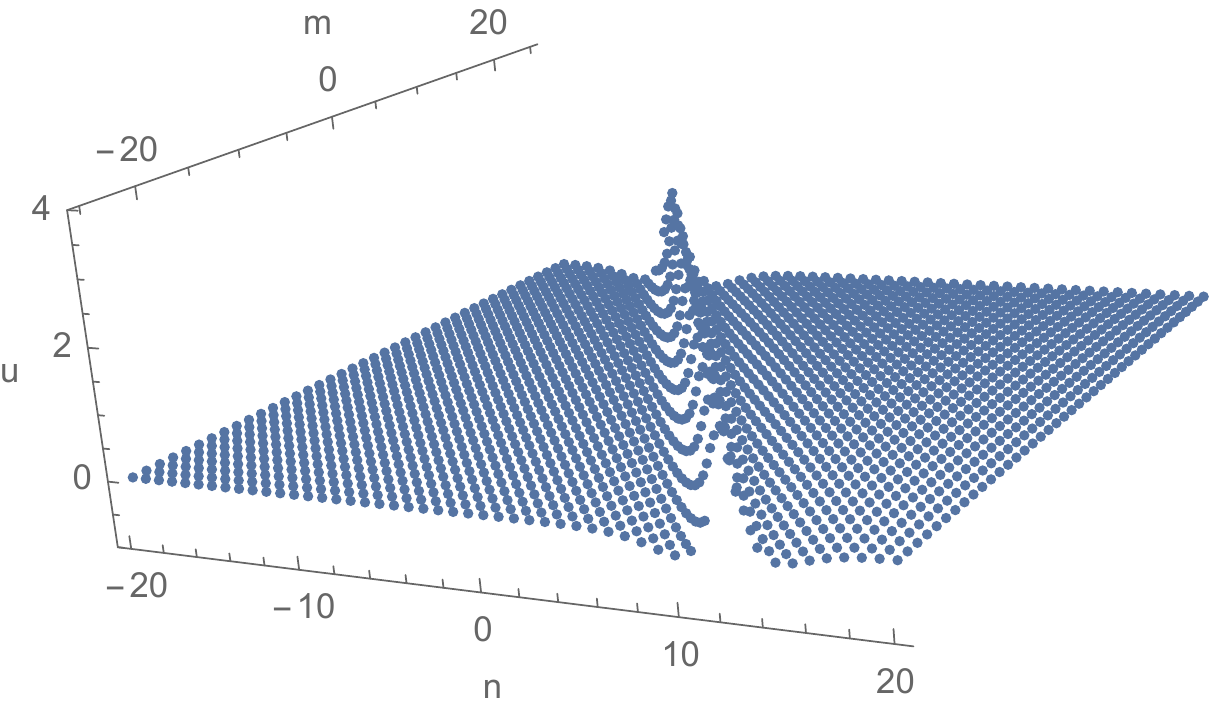}}}
		\put(80,-23){\resizebox{!}{4.0cm}{\includegraphics{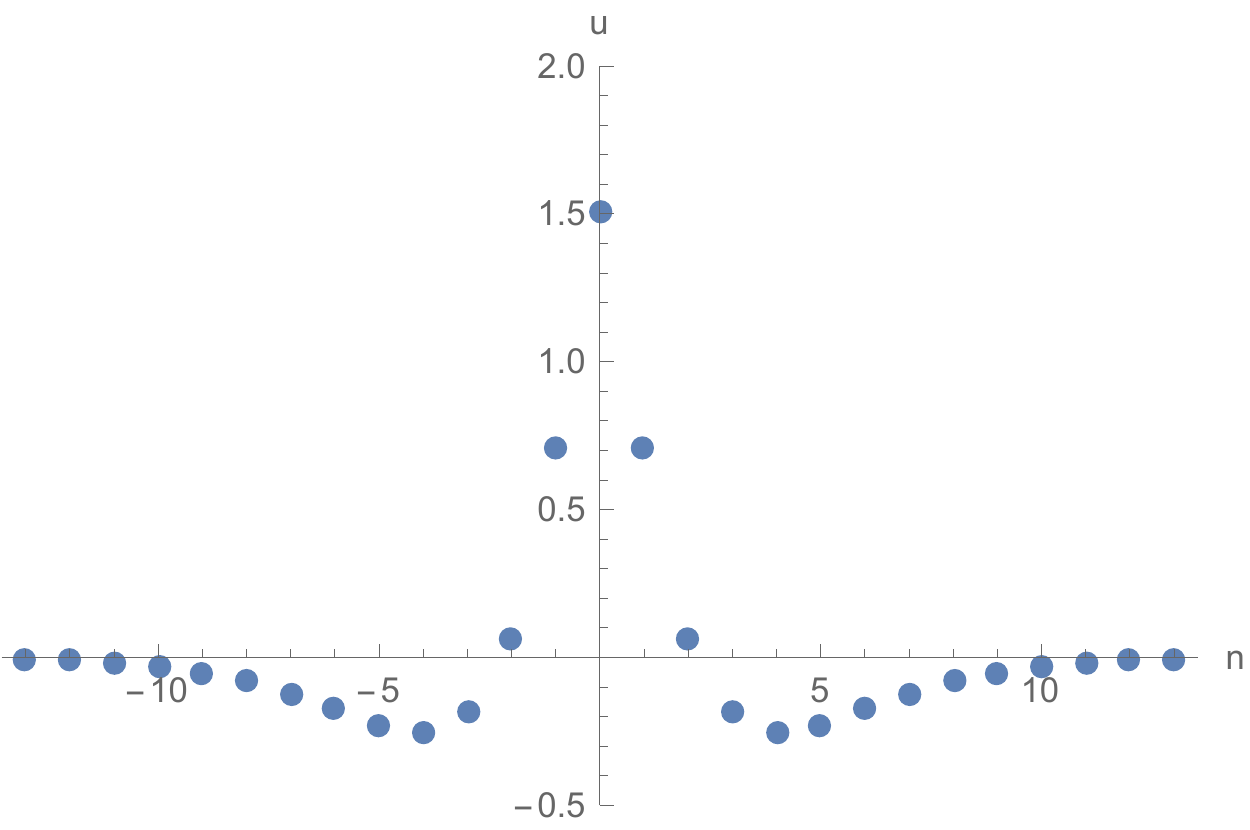}}}
	\end{picture}
\end{center}
\vskip 20pt
\begin{center}
	\begin{minipage}{15cm}{\footnotesize
			\qquad\qquad\qquad\qquad\quad(a)\qquad\qquad\qquad\qquad\qquad\qquad\qquad\qquad\qquad\qquad\qquad\quad (b) \\
			{\bf Fig. 3.} Jordan block solution $u$ given by \eqref{dr-JB-solu-u} with $k_1=0.3$ and $\delta=-2$:
			(a) shape and motion; (b) 2D-plot of (a) at $m=2$.}
	\end{minipage}
\end{center}

\subsection{Continuum limits}
\label{re-CL}

In the previous discussion, the real nonlocal discrete mKdV equation together with its soliton
solutions as well as Jordan block solutions has been derived. From the rnd-mKdV equation \eqref{rndmKdV},
through appropriate semi-continuous limit, one can obtain the real nonlocal semi-discrete
mKdV equation, i.e., differential-difference equation with one discrete and one continuous
independent variable. Furthermore, full continuous limit will be applied to yield
the real continuous nonlocal mKdV equation. Analogously, we denote these two equations by
rnsd-mKdV and rnc-mKdV, respectively. Here we will firstly use the indicative formula
\begin{align}
\label{dc-re}
\lim\limits_{m\rightarrow\infty}(1+k/m)^m=e^k
\end{align}
to consider continuum limits on the level of the discrete exponential function.
Then we will inspect the effect of the continuum limits to the equation themselves.
To this end, we concentrate on the discrete exponential function
\begin{align}
\label{e-xi}
e^{\xi}:=e^{kn}\bigg(\dfrac{\delta e^{2k}-1}{\delta e^{-2k}-1}\bigg)^{\frac{m}{2}}.
\end{align}

\vspace{.2cm}
\noindent{\bf Semi-continuous limit:} We start to perform the continuum limit by rewriting \eqref{e-xi} as
\begin{align}
\label{rstl-m}
\bigg(\dfrac{\delta e^{2k}-1}{\delta e^{-2k}-1}\bigg)^{\frac{m}{2}}
=\bigg(1+\dfrac{2\delta \sinh 2k}{\delta e^{-2k}-1}\bigg)^{\frac{m}{2}},
\end{align}
and taking the limit
\begin{align}
\label{rcl-nt}
\delta\rightarrow 0, \quad m\rightarrow\infty, \quad \text{s.t.}, \quad m\delta\rightarrow z.
\end{align}
The discrete exponential function \eqref{e-xi} gives rise to
\begin{align}
\label{lambdaj-def}
e^{\xi} \rightarrow e^{\lambda}, \quad \text{with} \quad \lambda:=kn-\sinh 2kz,
\end{align}
and the dependent variable $u(n,m)$ is reinterpreted as $\mu(n,z)$.
Inserting the Taylor expansions
\begin{subequations}
\label{Tay-1}
\begin{align}
& \wh{u}=\mu(z+\delta)=\mu+\delta\mu_{z}+\ldots, \\
& \wh{\dt{u}}=\dt{\mu}(z+\delta)=\dt{\mu}+\delta\dt{\mu}_{z}+\ldots,
\end{align}
\end{subequations}
into the rnd-mKdV equation \eqref{rndmKdV} and noting that $w\rightarrow 1$,
we realize that coefficient of the leading order $\mathcal{O}(\delta)$ is exactly
the rnsd-mKdV equation
\begin{align}	
\label{rnsd-z}
\mu_{z}+(1+\eta\mu\mu_{\sigma})(\wt\mu-\dt\mu)=0.
\end{align}
Similar to the discrete case,
equation \eqref{rnsd-z} is preserved under transformation $\mu\rightarrow -\mu$. And equation \eqref{rnsd-z} with
$(\sigma,\eta)=(\pm 1,1)$ and with $(\sigma,\eta)=(\pm 1,-1)$ can be
transformed into each other by taking $\mu\rightarrow i\mu$.

For the nonlocal equation \eqref{rnsd-z} with $\sigma=-1$, since its soliton and Jordan block solutions have been
studied well in Ref. \cite{FZS-IJMPB}, here we just give the following theorem and skip the detailed discussion.
\begin{Thm}
\label{Thm-rnsd}
The function $\mu=g/f$ with
\begin{align}
\label{rnsdsG-t-solu}
f=|e^{-N\Ga}\wh{\Phi^{(N)}};e^{N\Ga}T\wh{\Phi_{\sigma}^{(-N)}}|,\quad
g=|e^{-N\Ga}\widehat{\Phi^{(N+1)}};e^{N\Ga}T\wh{\Phi_{\sigma}^{(-N+1)}}|
\end{align}
solves the rnsd-mKdV equation \eqref{rnsd-z}, where $\Phi=e^{\Ga n-\sinh2\Ga z}C^{+}$ and $T$ is a constant matrix of order $2(N+1)$ satisfying
\begin{align}
\label{rnsdsG-t-Ga-T}
\Ga T+\sigma T\Ga=\bm 0,\quad T^{2}=\left\{
\begin{array}{l}
-\eta I, \quad \mbox{with} \quad \sigma=1,\\
\eta|e^\Gamma|^{2}I, \quad \mbox{with} \quad \sigma=-1.
\end{array}\right.
\end{align}
\end{Thm}

\noindent{\bf Full continuous limit:} We now pay attention to the
semi-discrete function $\lambda(z)$, which can be expanded as
\begin{subequations}
\label{c-lim}
\begin{align}
\lambda=(n-2z)k-4k^3z/3+\ldots.
\end{align}
A change of variable $(n-2z)\epsilon=x,~\epsilon^3z/3=t$ and $\kappa=k/\epsilon$ (see equation (1.9) in \cite{ZZW-JMP}) results into
\begin{align}
e^{\lambda}\rightarrow e^{\kappa x-4\kappa^3t}.
\end{align}
\end{subequations}
For the equation \eqref{rnsd-z}, we reinterpret the variable $\mu$ as $\mu(n,z):=\epsilon\alpha(x,t)/\sqrt{2}$.
Inserting the Taylor expansions
\begin{subequations}
\label{Tay-2}
\begin{align}	
&\mu_z=\frac{\epsilon^{2}}{\sqrt{2}}(-2\alpha_{x}+\frac{\epsilon^{2}}{3}\alpha_t), \\
&\wt{\mu}=\frac{\epsilon}{\sqrt{2}}\alpha(x+\epsilon)=\frac{\epsilon}{\sqrt{2}}
(\alpha+\epsilon\alpha_{x}+\dfrac{\epsilon^{2}}{2!}\alpha_{xx}+\dfrac{\epsilon^{3}}{3!}\alpha_{xxx}+\dots), \\
&\dt{\mu}=\frac{\epsilon}{\sqrt{2}}\alpha(x-\epsilon)=\frac{\epsilon}{\sqrt{2}}
(\alpha-\epsilon\alpha_{x}+\dfrac{\epsilon^{2}}{2!}\alpha_{xx}-\dfrac{\epsilon^{3}}{3!}\alpha_{xxx}+\dots)
\end{align}
\end{subequations}
into semi-discrete equation \eqref{rnsd-z}, we obtain as coefficient of the leading term the rnc-mKdV equation
\begin{align}
\label{rncmKdV}
\alpha_t+\alpha_{xxx}+3\eta\alpha\alpha_{\sigma}\alpha_{x}=0.
\end{align}
Naturally, equation \eqref{rncmKdV} is preserved under transformation $\al\rightarrow -\al$. And equation \eqref{rncmKdV} with
$(\sigma,\eta)=(\pm 1,1)$ and with $(\sigma,\eta)=(\pm 1,-1)$ can be
transformed into each other by taking $\al\rightarrow i\al$.

To proceed, let us consider the basic column vector $\Phi(n,z)$ appearing in Theorem \ref{Thm-rnsd}.
We absorb $e^{-N\Ga}$ into $\Phi(n,z)$ and redefine $\Phi(n,z):=\phi(x,t)$. Taking the Taylor expansions
\begin{align*}
E^{2j}\Phi(n,z)=\phi(x+2j\epsilon,t)=\phi+2j\epsilon\partial_{x}\phi+
\dfrac{(2j\epsilon)^{2}}{2!}\partial_{x}^{2}\phi+\ldots
\end{align*}	
with $j=1,2,\ldots,N+1$ into double Casoratian \eqref{rnsdsG-t-solu}, we consequently know\footnote{For the two basic
column vectors $\phi:=\phi(x,t)$ and $\psi:=\psi(x,t)$, the determinant here is the double
Wronskian, which is of form $|\wh{\phi^{(N)}}; \wh{\psi^{(N)}}|=|\phi^{(0)}, \phi^{(1)},\ldots,\phi^{(M)};
\psi^{(0)}, \psi^{(1)},\ldots,\psi^{(M)}|$, where $\phi^{(j)}=\frac{\partial^j\phi}{\partial x^j}$ and $\psi^{(j)}=\frac{\partial^j\psi}{\partial x^j}$.}
\begin{align}	
f\rightarrow |\Delta_{N}^{2}||\wh{\phi^{(N)}};T\wh{\phi_{\sigma}^{(N)}}|, \quad
g\rightarrow |\Delta_{N+1}\Delta_{N-1}||\wh{\phi^{(N+1)}};T\wh{\phi_{\sigma}^{(N-1)}}|,
\end{align}	
where
\begin{align}
	\Delta_{N}=
	\left(
	\begin{array}{cccc}
		2\epsilon &  \dfrac{(2\epsilon)^{2}}{2!} &  \cdots& \dfrac{(2\epsilon)^{N}}{N!}\\
		4\epsilon &  \dfrac{(4\epsilon)^{2}}{2!} &  \cdots& \dfrac{(4\epsilon)^{N}}{N!}\\		
		\vdots& \vdots&\cdots&\vdots\\
		2N\epsilon&\dfrac{(2N\epsilon)^{2}}{2!}& \cdots & \dfrac{(2N\epsilon)^{N}}{N!}
	\end{array}\right).
\end{align}
Therefore, for the dependent variable $\mu$ we have
\begin{align}
\label{u-al-lim}
\mu=\dfrac{g}{f}\rightarrow
\frac{\epsilon}{\sqrt{2}}\al=\dfrac{2\epsilon|\wh{\phi^{(N+1)}};T\wh{\phi_{\sigma}^{(N-1)}}|}
{|\wh{\phi^{(N)}};T\wh{\phi_{\sigma}^{(N)}}|}.
\end{align}
To summarize, we end this section by the following theorem,
which was firstly given in \cite{CDLZ}.
\begin{Thm}
\label{Thm-rnc}
The function $\al=g/f$ with
\begin{align}
\label{rnl-solu}
f=|\wh{\phi^{(N)}};T\wh{\phi_{\sigma}^{(N)}}|,\quad
g=2\sqrt{2}|\widehat{\phi^{(N+1)}};T\wh{\phi_{\sigma}^{(N-1)}}|,
\end{align}
solves the equation \eqref{rncmKdV}, where $\phi=e^{\Ga x-4\Ga^3t}C^{+}$ and $T$ is a constant matrix of order $2(N+1)$ satisfying
\begin{align}
\label{rnl-c-Ga-T}
\Ga T+\sigma T\Ga=\bm 0,\quad T^{2}=-\eta\sigma I.
\end{align}
\end{Thm}

\section{Complex reductions: solutions and continuum limits}

The dAKNS equation \eqref{dAKNS} admits complex local and nonlocal reduction
\begin{align}
\label{com-Re}
v=\eta u^*_\sigma, \quad w=w^*_\sigma, \quad \eta, \sigma=\pm 1,
\end{align}
under which the reduced equation is
\begin{subequations}
\label{cnd-mKdV}
\begin{align}
& \widehat{u}-u+\delta(1+\eta u u^{*}_\sigma)(\widetilde{u}-\widehat{\underaccent{\tilde}{u}})w=0, \\
\label{cn-mKdV-rb}
& (1+\eta\wh{u}\wh{u}^{*}_{\sigma})\wt w=(1+\eta uu^{*}_{\sigma})w.
\end{align}
\end{subequations}
When $\sigma=1$, \eqref{cnd-mKdV} is a local equation, while when
$\sigma=-1$, \eqref{cnd-mKdV} is a nonlocal equation. This equation
is preserved under transformations $u\rightarrow -u$ and $u\rightarrow \pm iu$.
For convenience, we name \eqref{cnd-mKdV} as cnd-mKdV equation.

For the double Casoratian solutions to cnd-mKdV equation \eqref{cnd-mKdV}, we have the following result.
\begin{Thm}
The functions $u=g/f$ and $w=f\wh{\dt f}/(\wh{f}\dt{f})$ with
\begin{align}
\label{cndmKdV-solu}
f=|e^{-N\Ga}\wh{\Phi^{(N)}};e^{N\Ga}T\wh{\Phi_{\sigma}^{*(-N)}}|,\quad
g=|e^{-N\Ga}\wh{\Phi^{(N+1)}};e^{N\Ga}T\wh{\Phi_{\sigma}^{*(-N+1)}}|
\end{align}
solve the cnd-mKdV equation \eqref{cnd-mKdV}, if the $(2N+2)$-th order column vector	$\Phi$ is
defined as \eqref{dAKNS-Phi-Ga}, where $T\in\mathbb{C}^{(2N+2)\times(2N+2)}$ is a constant matrix satisfying
\begin{align}
\label{cndmKdv-AT}
\Ga T+\sigma T\Ga^*=\bm 0,\quad TT^*=\left\{
\begin{array}{l}
-\eta I, \quad \mbox{with} \quad \sigma=1,\\
\eta|e^{\Ga^*}|^{2}I, \quad \mbox{with} \quad \sigma=-1.
\end{array}\right.
\end{align}
\end{Thm}

\subsection{One-soliton solution}

Because the complex conjugate is involved, in what follows we just address the
one-soliton solution for equation \eqref{cnd-mKdV} with $(\eta,\sigma)=(1,-1)$.
In this case, we take
\begin{align}
\label{cd-LT-form}
\Ga=\left(
\begin{array}{cc}
L & \bm 0  \\
\bm 0 & L^*
\end{array}\right),
\quad
T=\left(
\begin{array}{cc}
\bm 0 & I  \\
I & \bm0
\end{array}\right)|e^{\Ga^{*}}|,
\end{align}	
where $L$ is the diagonal matrix \eqref{drL-Diag} with different complex discrete
spectral parameters $\{k_j\}$. Then the basic column vector $\Phi$ is composed of
\begin{align}
\label{no}
\Phi_{j}=\left\{
\begin{aligned}
& c_{j}e^{\xi_j}, \quad j=1, 2, \ldots, N+1, \\
& d_{s}e^{\xi_s^*}, \quad j=N+1+s,\quad s=1,2,\dots,N+1,
\end{aligned}	
\right.
\end{align}
where and whereafter $\{c_{j},d_{s}\}$ are complex constants.

When $N=0$, the one-soliton solution reads
\begin{subequations}
\begin{align}
& \label{cnl-solu}
u=\dfrac{2c_{1}d_{1}\sinh(k_{1}^{*}-k_{1})}
{|c_{1}|^{2}e^{-2\xi_{1}^{*}}-|d_{1}|^{2}e^{-2\xi_{1}}}, \\
& w=f\wh{\dt f}/(\wh{f}\dt{f}), \quad f=|c_{1}|^{2}e^{\xi_1-\xi_1^{*}}-|d_{1}|^{2}e^{\xi_1^*-\xi_1},
\end{align}
\end{subequations}
where $\xi_1$ is defined by \eqref{xi}. To demonstrate the dynamics of solution $u$ in an analytic way, we
take $k_{1}=k_{11}+ik_{12}$ and write \eqref{cnl-solu} as following form
\begin{subequations}
\begin{align}
\label{dc-1ss}
u=& \dfrac{-2ic_{1}d_{1}\varepsilon_1^{m/2}e^{2nk_{11}}\sin2k_{12}}
{|c_{1}|^{2} e^{i(2nk_{12}+m\varepsilon_2)}	-|d_{1}|^{2}e^{-i(2nk_{12}+m\varepsilon_2)}}, \\
\varepsilon_1=& \dfrac{(\delta^{2}\sin4k_{12}-2\delta\cosh2k_{11}\sin2k_{12})^{2}+(\delta^{2}\cos4k_{12}-2\delta\cosh2k_{11}\cos2k_{12}+1)^{2}}
{(\delta^{2}e^{-4k_{11}}-2\delta e^{-2k_{11}}\cos2k_{12}+1)^{2}}, \\
\varepsilon_2=& \arctan\dfrac{\delta^{2}\sin4k_{12}-2\delta\cosh2k_{11}\sin2k_{12}}
{\delta^{2}\cos4k_{12}-2\delta\cosh2k_{11}\cos2k_{12}+1}.
\end{align}	
\end{subequations}
After straightforward manipulations one arrives at
\begin{align}
	\label{dc-1ss-mo}
	|u|^2=\dfrac{4|c_1d_1|^2\varepsilon_1^{m}e^{4nk_{11}}\sin^22k_{12}}
	{|c_1|^{4}+|d_{1}|^{4}
		-2|c_1d_1|^2\cos(4nk_{12}+2m\varepsilon_2)},
\end{align}
which is oscillatory due to the cosine in the denominator.
When $|c_{1}|=|d_{1}|$, it is evident that solution \eqref{dc-1ss-mo} has singularities along points
\begin{align}
n(m)=\dfrac{\ell\pi-m\varepsilon_2}{2k_{12}},\quad \ell\in\mathbb{Z}.
\end{align}
While when $|c_{1}|\neq|d_{1}|$, solution \eqref{dc-1ss-mo} is nonsingular travelling wave with
velocity $-\varepsilon_{2}/(2k_{12})$, and reaches its extrema along points
\begin{align}
	n(m)=-\dfrac{m\varepsilon_2}{2k_{12}}+\frac{1}{4k_{12}}\bigg(\arcsin\frac{(|c_1|^4+|d_1|^4)k_{11}}
	{2|c_1d_1|^2\sqrt{k^2_{11}+k^2_{12}}}-\jmath+2\ell\pi\bigg),\quad \ell\in\mathbb{Z},
\end{align}
where $\sin\jmath=k_{11}/\sqrt{k^2_{11}+k^2_{12}}$.
For any $m$, $|u|^2$ goes to zero as either $(k_{11}<0,n\rightarrow +\infty)$ or $(k_{11}>0,n\rightarrow -\infty)$.
\vskip20pt
\begin{center}
\begin{picture}(120,80)
\put(-160,-23){\resizebox{!}{4.0cm}{\includegraphics{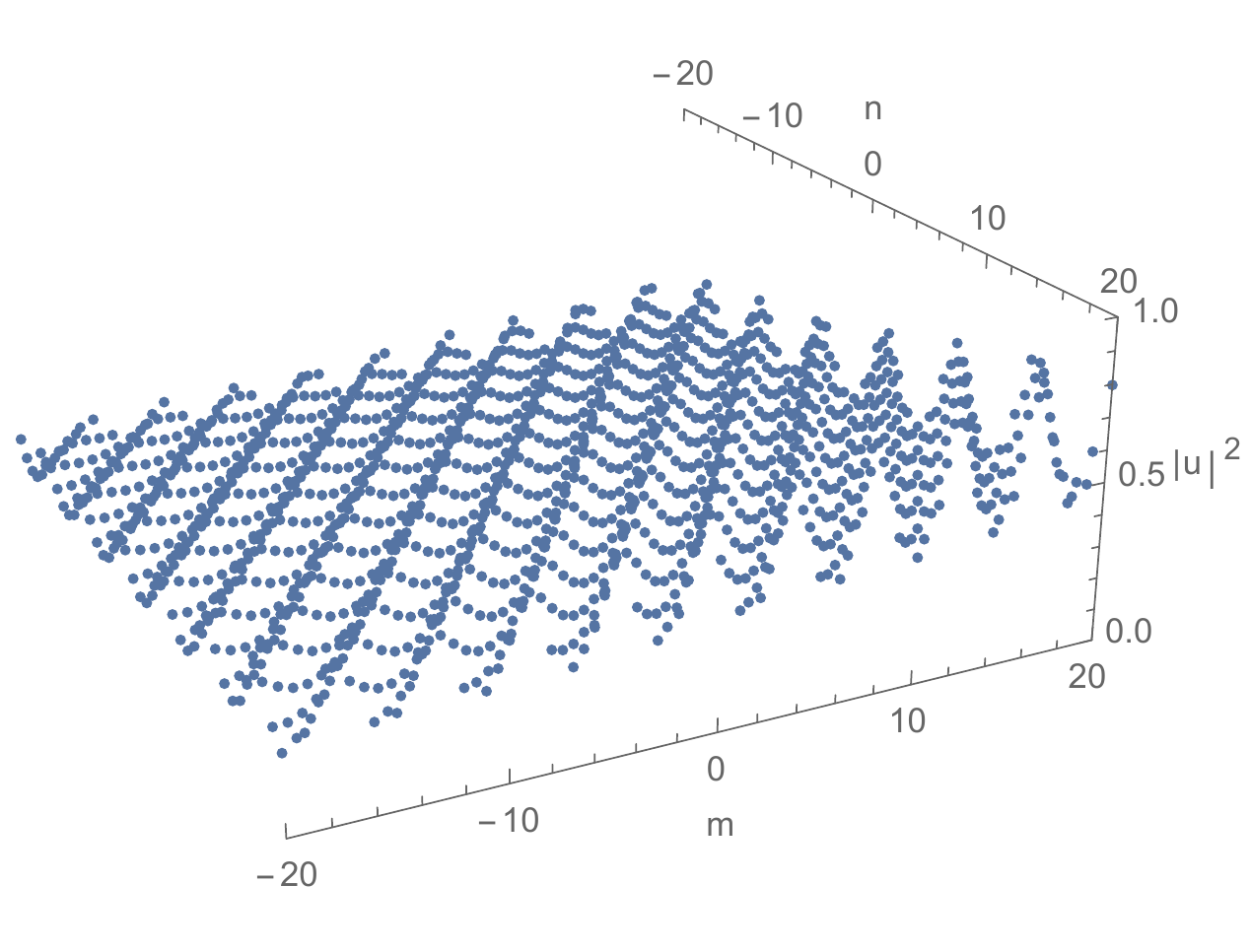}}}
\put(0,-23){\resizebox{!}{3.5cm}{\includegraphics{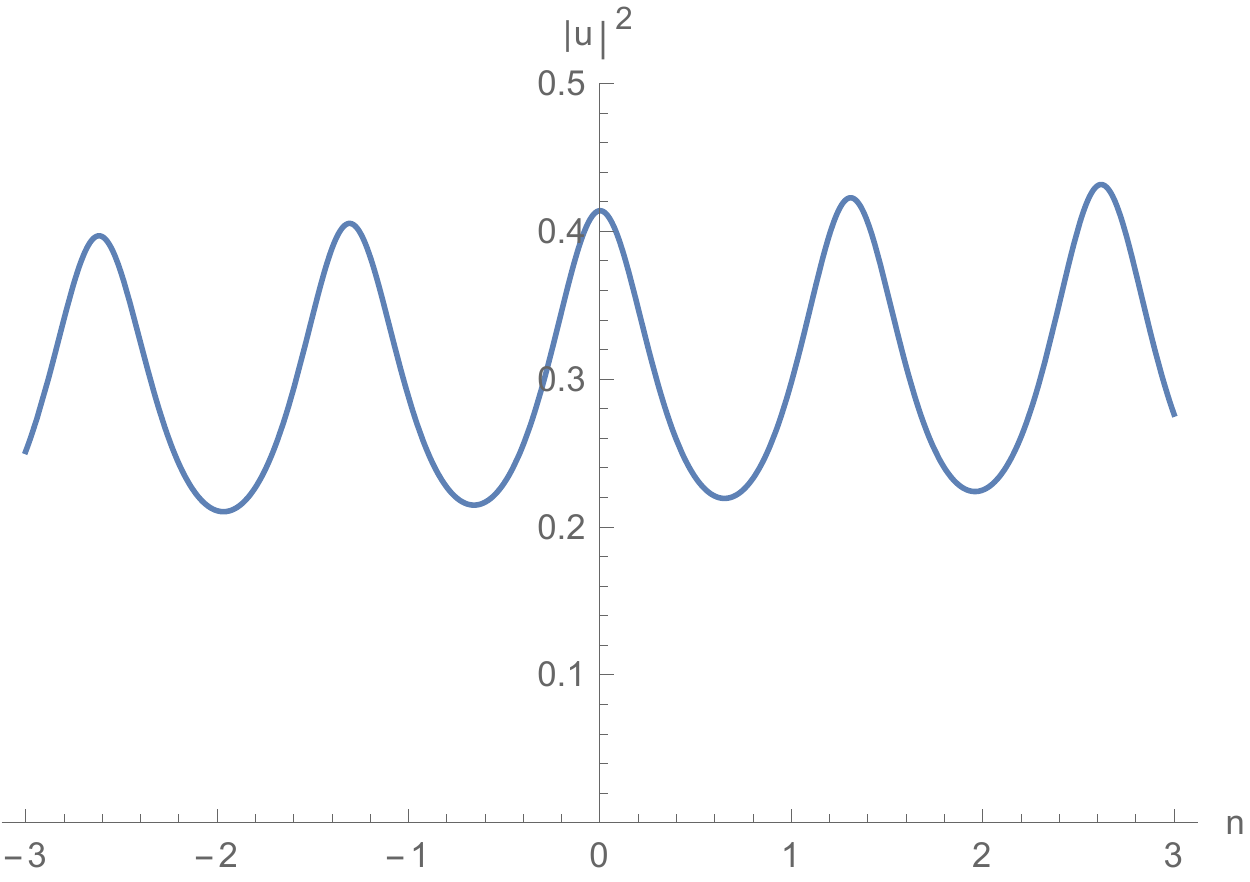}}}
\put(150,-23){\resizebox{!}{3.5cm}{\includegraphics{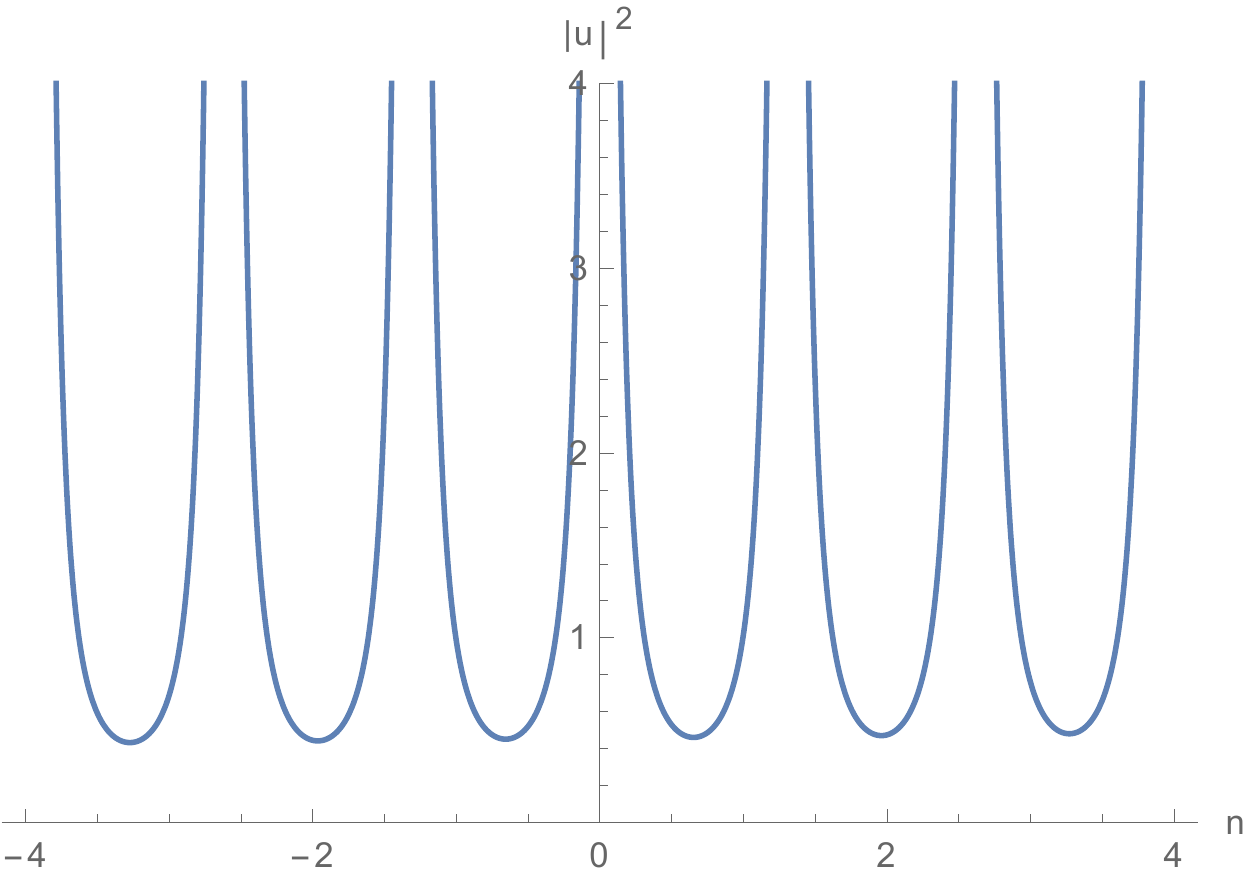}}}
\end{picture}
\end{center}
\vskip 20pt
\begin{center}
\begin{minipage}{15cm}{\footnotesize
\qquad\qquad\qquad\quad(a)\quad\qquad\qquad\qquad\qquad\qquad\qquad\qquad (b) \quad\qquad\qquad\qquad\qquad\qquad\qquad\quad (c)\\
{\bf Fig. 4}. One-soliton solution $|u|^2$ given by \eqref{dc-1ss-mo} with $\delta=-1.1$ and $k_1=0.004+1.2i$:
(a) shape and movement with $c_1=0.5+0.5i$ and $d_1=0.2+0.2i$;
(b) 2D-plot of (a) at $m=0$;
(c) 2D-plot with $c_1=d_1=1+i$ at $m=0$.}
\end{minipage}
\end{center}

\subsection{Continuum limits} The continuum limits adopted here are the same as the real case. In other words,
we can still apply semi-continuous limit \eqref{rcl-nt}
to obtain the complex local and nonlocal semi-discrete mKdV equation, and continuous limit \eqref{c-lim}
to derive the complex local and nonlocal continuous mKdV equation. For the sake of simplicity,
we denote the two resulting equations by cnsd-mKdV and cnc-mKdV.

\vspace{.2cm}
\noindent{\bf Semi-continuous limit:} Under the semi-continuous limit \eqref{rcl-nt},
inserting the Taylor expansions \eqref{Tay-1} into the cnd-mKdV equation \eqref{cnd-mKdV} and using $w\rightarrow 1$,
we obtain as coefficient of the leading term of order $\mathcal{O}(\delta)$ the cnsd-mKdV equation
\begin{align}	
\label{cnsdmKdV-z}
\mu_z+(1+\eta\mu\mu_{\sigma}^{*})(\wt\mu-\dt\mu)=0.
\end{align}
This equation is preserved under transformations $\mu\rightarrow -\mu$ and $\mu\rightarrow \pm i\mu$.
For the double Casoratian solutions to equation \eqref{cnsdmKdV-z}, we have the following result.
\begin{Thm}
The function $\mu=g/f$ with
\begin{align}
\label{cnsdmKdV-t-solu}
f=|e^{-N\Ga}\wh{\Phi^{(N)}};e^{N\Ga}T\wh{\Phi_{\sigma}^{*(-N)}}|,\quad
g=|e^{-N\Ga}\widehat{\Phi^{(N+1)}};e^{N\Gamma}T\wh{\Phi_{\sigma}^{*(-N+1)}}|,
\end{align}
solves the cnsd-mKdV equation \eqref{cnsdmKdV-z}, where $\Phi=e^{\Ga n-\sinh 2\Ga t}C^{+}$ and $T$ is a constant matrix of order $2(N+1)$
satisfying
\begin{align}
\label{cnsdsG-t-Ga-T}
\Ga T+\sigma T\Ga^*=\bm 0,\quad TT^*=\left\{
\begin{array}{l}
-\eta I, \quad \mbox{with} \quad \sigma=1,\\
\eta|e^{\Ga^*}|^{2}I, \quad \mbox{with} \quad \sigma=-1.
\end{array}\right.
\end{align}
\end{Thm}

\vspace{.2cm}
\noindent{\bf Full continuous limit:} Employing the full continuous limit \eqref{c-lim} and Taylor expansions
\eqref{Tay-2} to the cnsd-mKdV equation \eqref{cnsdmKdV-z}, we further arrive at the cnc-mKdV equation
\begin{align}	
\label{cncmKdV}
\alpha_t+\alpha_{xxx}+3\eta\alpha\alpha_{\sigma}^{*}\alpha_{x}=0,
\end{align}	
which is preserved under transformations $\al\rightarrow -\al$ and $\al\rightarrow \pm i\al$.

Double Casoratian solutions to the cnc-mKdV equation \eqref{cncmKdV} can be summarized by the following theorem.
\begin{Thm}
\label{Thm-cncmKdV-solu}
The function $\al=g/f$ with
\begin{align}
\label{cncmKdV-solu}
f=|\wh{\phi^{(N)}};T\wh{\phi_{\sigma}^{*(N)}}|,\quad
g=2\sqrt{2}|\wh{\phi^{(N+1)}};T\wh{\phi_{\sigma}^{*(N-1)}}|,
\end{align}
solves the cnc-mKdV equation \eqref{cncmKdV}, where $\phi=e^{\Ga x-4\Ga^{3} t}C^{+}$ and $T$ is a constant matrix of order $2(N+1)$ satisfying
\begin{align}
\label{cncsG-B-T}
\Gamma T+\sigma T\Gamma^{*}=\bm 0,\quad TT^*=-\eta\sigma I.
\end{align}
\end{Thm}

\section{Conclusions}

In this paper, we considered double Casoratian solutions of the dAKNS equation \eqref{dAKNS}. By imposing real/complex
local and nonlocal reduction on the two dependent variables of the dAKNS equation, we derived
real/complex local and nonlocal discrete mKdV equations \eqref{rndmKdV} and \eqref{cnd-mKdV}.
To the rnd-mKdV equation \eqref{rndmKdV}, we investigated its one-soliton,
two-soliton and the simplest Jordan block solutions. Dynamics of one-soliton and
two-soliton solutions were analyzed and illustrated. Applying the semi-continuous limit,
as well as full continuous limit, we recover the obtained discrete mKdV equation to the
semi-discrete mKdV equation and continuous mKdV equation.
For the cnd-mKdV equation \eqref{cnd-mKdV}, we mainly discussed its one-soliton solution and
dynamics. The same semi-continuous limit and full continuous limit were employed to
obtain the semi-discrete and continuous counterparts of cnd-mKdV equation \eqref{cnd-mKdV}.
On account of the involvement of cosine function in soliton solutions to complex nonlocal mKdV equations, regardless of
discrete, semi-discrete or continuous, the solutions behave quasi-periodically.

\section*{Acknowledgments}

This project is supported by the National Natural Science Foundation of
China (No. 12071432) and the Natural Science Foundation of Zhejiang Province (No. LY17A010024).

\end{document}